\documentclass[aps,prb,twocolumn,showpacs,amssymb,superscriptaddress]{revtex4}
\usepackage{amsmath,bm}
\usepackage{graphicx,subfigure}

\begin{document}

\title{Scattering universality classes of side jump in anomalous Hall effect}

\author{Shengyuan A. Yang}
\affiliation{Department of Physics, The University of Texas, Austin,
Texas 78712, USA}

\author{Hui Pan}
\affiliation{Department of Physics, The University of Texas, Austin,
Texas 78712, USA} \affiliation{Department of Physics, Beijing
University of Aeronautics and Astronautics, Beijing 100083, China}

\author{Yugui Yao}
\affiliation{Department of Physics, The University of Texas, Austin,
Texas 78712, USA} \affiliation{Beijing National Laboratory for
Condensed Matter Physics and Institute of Physics, Chinese Academy
of Sciences, Beijing 100190, China}

\author{Qian Niu}
\affiliation{Department of Physics, The University of Texas, Austin,
Texas 78712, USA}

\date{\today}


\begin{abstract}
The anomalous Hall conductivity has an important extrinsic
contribution known as side jump contribution, which is independent
of both scattering strength and disorder density. Nevertheless, we
discover that side jump has strong dependence on the spin structure
of the scattering potential. We propose three universality classes
of scattering for the side jump contribution, having the characters
of being spin-independent, spin-conserving and spin-flip
respectively. For each individual class, the side jump contribution
takes a different unique value. When two or more classes of
scattering are present, the value of side jump is no longer fixed
but varies as a function of their relative disorder strength. As
system control parameter such as temperature changes, due to the
competition between different classes of disorder scattering, the
side jump Hall conductivity could flow from one class dominated
limit to another class dominated limit. Our result indicates that
magnon scattering plays a role distinct from normal impurity
scattering and phonon scattering in the anomalous Hall effect
because they belong to different scattering classes.
\end{abstract}

\maketitle

\section{INTRODUCTION}
The anomalous Hall effect (AHE), in which a transverse voltage is
induced by a longitudinal current flow in ferromagnetic materials,
is one of the most intriguing effects in condensed matter physics.
While it has been widely used experimentally as a standard technique
for the characterization of ferromagnet materials, the theoretical
formulation of AHE proves to be complicated and is a subject full of
controversial issues and conflicting results.~\cite{naga2009} In
recent years, an important connection has been established between
AHE and the Berry phase of Bloch
electrons.~\cite{berr1984,sund1999,jung2002,onod2002} This triggers
revived interest in this subject and is followed by extensive
researches both theoretically and
experimentally.\cite{ye1999,tagu2001,crep2001,fang2003,culc2003,schl2003,sino2004,yao2004,hald2004,lee2004,
sini2005,duga2005,kotz2005,zeng2006,liu2006,inou2006,onod2006,yao2007,sini2007prb,boru2007,kato2007,nunn2007,liu2007,
sini2008,onod2008,nunn2008,rash2008,venk2008,arse2008,kova2009,sang2009,shio2009,glun2009,tian2009,sino2010,dimi2010}

It is now generally accepted that spin-orbit coupling and spin
splitting are two essential ingredients for AHE, and apart from an
intrinsic contribution which is scattering independent, there are
also important extrinsic contributions to AHE due to disorder
scattering. Based on their parametric dependence on the disorder
density $n_\text{dis}$, the extrinsic contributions can be collected
into two subgroups: the side jump contribution of order
$n_\text{dis}^0$ and the skew scattering contribution of order
$n_\text{dis}^{-1}$.

The side jump contribution is of special interest in that it arises
from scattering, but surprisingly it does not depend on either the
scattering strength or the disorder density (we shall use the term
``disorder strength" to stand for both scattering strength and
disorder density). Theoretical calculations of simple model systems
show that the side jump contribution is usually at least as
important as intrinsic
contribution.\cite{sini2005,duga2005,onod2006,sini2007prb} However,
the good agreement between the intrinsic contribution calculated
from first principles and the experimental results seems to indicate
that the side jump contribution only plays a subdominant
role.\cite{yao2004,zeng2006} This remains as a puzzle that need to
be resolved and is partly our motivation for the present work.

Historically, the concept of side jump was first devised by
Berger,\cite{berg1970} which refers to the coordinate shift of a
wave-packet during an impurity scattering and this process leads to
a contribution of order $n_\text{dis}^0$ to the anomalous Hall
conductivity. Recently, it has been found that besides this
coordinate shift process, several other scattering processes also
generate contributions of order $n_\text{dis}^0$.\cite{sini2007prb}
It should be noted that the term ``side jump" used in the present
paper includes all the scattering induced contributions of order
$n_\text{dis}^0$, not only the contribution from Berger's original
side jump.

In the study of physical systems, properties which are insensitive
to detailed parameter values and system configurations but are only
determined by the symmetry are especially interesting and important.
It is helpful to define universality classes based on the behavior
of these universal properties under certain imposed symmetry and
study the generic properties of each class. Side jump can be
regarded as a universal property for a disordered system in the
sense that its value does not depend on the detailed disorder
profile, but we shall see that it has sensitive dependence on the
symmetry property of the scattering. Consequently, it is natural to
define universality classes of disorder scattering according to
their side jump contributions and study the anomalous Hall response
for each class.

In this work, we propose three universality classes of disorder
scattering, each has different structures in spin space. We find
that: (1) for each individual class, the side jump contribution
takes a distinct value independent of the detailed disorder profile.
In particular, we show that magnon scattering plays a distinct role
from both impurity scattering and phonon scattering in AHE; (2) when
several classes of scattering are present, side jump depends on
their relative disorder strength and a sign change is possible as a
result of their competition. Since in real physical system
scattering processes of all the three classes exist, our finding
indicates that a careful classification and analysis of different
scattering processes is indispensible for an accurate account of
AHE.

This paper is organized as the following. First, in Sec. II, we
propose and discuss the three scattering universality classes of
side jump. In Sec. III we demonstrate our ideas by a concrete
analytical calculation of anomalous Hall conductivity of massive
Dirac model for each universality class. In Sec. IV we discuss the
important consequences of our result, especially about contribution
from magnon scattering in AHE, and draw some final conclusions.

\section{UNIVERSALITY CLASSES OF DISORDER SCATTERING}
The general form of a random disorder potential for carriers with
spin (or pseudospin) degrees of freedom can be written as
\begin{equation}
\hat{V}_\text{dis}(\bm{r})=\sum_i\left[V_0(\bm{r}-\bm{R}_i)+\bm{V}(\bm{r}-\bm{R}_i)\cdot
\hat{\bm{\sigma}}\right],
\end{equation}
where $\bm{R}_i$ ($i=1,2,\cdots$) are the positions of randomly
distributed scattering centers,
$\hat{\bm{\sigma}}=(\hat{\sigma}_x,\hat{\sigma}_y,\hat{\sigma}_z)$
is a vector with components of Pauli matrices and the hat means the
quantity is a $2\times 2$ matrix in spin space. We assume that the
statistical average of the disorder potential is zero (any nonzero
value only shifts the origin of total energy) and the second order
spatial correlation only depends on the difference in positions,
\begin{equation}
\langle V_\text{dis}(\bm{r})\rangle_c=0,\qquad \langle
V_\text{dis}(\bm{r})V_\text{dis}(\bm{r}')\rangle_c=B(\bm{r}-\bm{r}'),
\end{equation}
where the angular bracket $\langle\cdots\rangle_c$ denotes disorder
average. In order to discuss skew scattering which originates from
higher order scattering processes, we allow a non-vanishing third
order disorder correlation instead of requiring the disorder to be
purely Gaussian.

Time reversal symmetry has to be broken for the appearance of
AHE.\cite{naga2009} In ferromagnet, this is realized by the
spontaneous magnetic ordering. We will be most interested in the
configuration that the magnetization is perpendicular to the
two-dimensional (2D) plane where the transport occurs, as is
pertinent for most experimental investigations. It is reasonable to
assume that over disorder average the system is isotropic in the 2D
plane with no preferred in-plane directions. With this symmetry
constraint, the total angular momentum (in the direction normal to
the plane which we will refer to as $z$-axis) of the carriers are
conserved on average. Due to spin-orbit coupling, the carrier's
orbital motion which is tied to orbital angular momentum depends
sensitively on the change of its spin angular momentum during a
scattering. Based on this consideration, we propose the following
three classes of disorder scattering which as we will see lead to
different values of side jump contribution:
\begin{equation}\begin{split}
&\text{Class A}\qquad \hat{V}=V^\text{o}\hat{1},\\
&\text{Class B}\qquad \hat{V}=V^\text{o}\hat{\sigma}_z,\\
&\text{Class C}\qquad \hat{V}=V^\text{o}\hat{\sigma}_\pm/\sqrt{2},\\
\end{split}
\end{equation}
where $V^\text{o}$ denotes the orbital part of the scattering
potential, and $\hat{\sigma}_\pm\equiv \hat{\sigma}_x\pm
i\hat{\sigma}_y$. Each class has a different action on the carrier's
spin. Class A is isotropic in spin space. Class B, like Class A,
conserve the $z$-component of the carrier spin but spin up and spin
down carriers experience different scattering potentials. Class C,
unlike the first two classes, induces spin flips. The three classes
as we discuss later represent a quite general classification scheme
for real physical systems. This classification scheme is also
evident if we consider the disorder correlation function under the
in-plane rotational symmetry,\cite{expl}
\begin{equation}
\begin{split}\label{Vcor}
\left\langle
V_\text{dis}^{ij}(\bm{r})V_\text{dis}^{ji}(\bm{r}')\right\rangle_c=\;&
\langle V_0V_0\rangle_c\delta_{ij}+ \langle
V_zV_z\rangle_c(\hat{\sigma}_z)_{ij}(\hat{\sigma}_z)_{ji}\\
&+\sum_{\alpha=x,y} \langle V_\alpha
V_\alpha\rangle_c(\hat{\sigma}_\alpha)_{ij}(\hat{\sigma}_\alpha)_{ji},
\end{split}
\end{equation}
where $i,j$ are spin indices. Since $\langle V_xV_x\rangle_c=\langle
V_yV_y\rangle_c$ due to the in-plane rotational symmetry, the last
term is proportional to
\begin{equation}\label{classC}
\left[(\hat{\sigma}_+)_{ij}(\hat{\sigma}_-)_{ji}+(\hat{\sigma}_-)_{ij}(\hat{\sigma}_+)_{ji}\right],
\end{equation}
with each term being invariant under spin rotations around $z$-axis.
The three terms in Eq.(\ref{Vcor}) just correspond to the three
classes we defined.

Before proceeding, we point out an important difference between
Class C and Class A, B on their third order correlation functions.
The Class C disorder can be expressed as
$\bm{V}(\bm{r})\cdot\bm{\sigma}$ where $\bm{V}$ is a random in-plane
vector. Under in-plane rotational symmetry, $\bm{V}$ has no
preferred direction therefore its third order correlation like
$\langle VVV\rangle_c$ must vanish. However for Class A or Class B,
the third order correlation is not dictated by this symmetry
constraint hence does not necessarily vanish. This difference will
be reflected in the skew scattering contribution to the AHE.

The transverse motion of carriers in AHE is a result of spin-orbit
coupling. In our classification scheme, each class of scattering has
different effects on the carrier spin, hence will also have
different effects on the carrier orbits. This is the underlying
reason for their distinct contributions to the AHE and especially
the side jump part. In the following section, we demonstrate this
idea by a concrete model calculation.

\section{AHE OF MASSIVE DIRAC MODEL}
\subsection{Model and Approach}
To demonstrate the rationale of our classification scheme, we
calculate the anomalous Hall conductivity for the massive Dirac
model. This model is usually considered as the minimal model for
AHE.\cite{naga2009} The model Hamiltonian reads (we set $\hbar=1$
and assume $\Delta>0$ in the following calculations)
\begin{equation}\label{dirac}
\hat{\mathcal{H}}=v(k_x\hat{\sigma}_x+k_y\hat{\sigma}_y)+\Delta
\hat{\sigma}_z,
\end{equation}
where spin-orbit coupling is contained in the first term with $v$
being the coupling constant, and the last term breaks the time
reversal symmetry and is also responsible for the finite electron
mass at the band edge. This model captures interesting physics near
a generic band anti-crossing point due to spin-orbit coupling.

The eigenstates of the system are
\begin{equation}
\psi_{\bm{k}}^\pm(\bm{r})=\frac{1}{\sqrt{A}}e^{i\bm{k}\cdot\bm{r}}|u_{\bm{k}}^\pm\rangle,
\end{equation}
with the corresponding energy eigenvalues
\begin{equation}\label{dispersion}
\varepsilon^\pm(\bm{k})=\pm\sqrt{(vk)^2+\Delta^2},
\end{equation}
where $\pm$ labels the upper and lower band respectively, $A$ is the
system size,  and $|u_{\bm{k}}^\pm\rangle$ is the spin part of the
eigenstate which can be written as
\begin{equation}
|u_{\bm{k}}^+\rangle=\left(
                       \begin{array}{c}
                         \cos\frac{\theta}{2} \\
                         \sin\frac{\theta}{2}e^{i\phi} \\
                       \end{array}
                     \right),
\qquad |u^-_{\bm{k}}\rangle=\left(
                       \begin{array}{c}
                         \sin\frac{\theta}{2} \\
                         -\cos\frac{\theta}{2}e^{i\phi} \\
                       \end{array}
                     \right),
\end{equation}
where $\theta$ and $\phi$ are the spherical angles of the vector
$(vk_x,vk_y,\Delta)$ such that
\begin{equation}\begin{split}
&\cos\theta=\frac{\Delta}{\sqrt{(vk)^2+\Delta^2}},\qquad
\sin\theta=\frac{vk}{\sqrt{(vk)^2+\Delta^2}},\\
&\tan\phi=k_y/k_x.
\end{split}
\end{equation}
Due to spin-orbit coupling, the spin state is a function of the
momentum $\bm{k}$. The energy spectrum consists of two anti-crossing
bands with a band gap of $2\Delta$. From the dispersion relation
Eq.(\ref{dispersion}), the geometry of the bands can be termed as a
``Dirac hyperboloid" (of two sheets).

To calculate the anomalous Hall conductivity, we follow Sinitsyn
\emph{et al.}\cite{sini2007prb} by using the Kubo-Streda
formalism.\cite{stre1982,crep2001} In this approach, the Hall
conductivity can be separated into two parts in the weak scattering
regime, $\sigma_{xy}=\sigma_{xy}^\text{I}+\sigma_{xy}^\text{II}$,
where $\sigma_{xy}^\text{I}$ is a Fermi surface contribution which
includes all the important scattering contributions, and
$\sigma_{xy}^\text{II}$ is a Fermi sea contribution for which we
only need to retain the scattering-free
component.~\cite{sini2007prb} In the following we consider that the
system is electron doped with Fermi energy $\varepsilon_F>\Delta$
and due to particle-hole symmetry, the results can be easily
generalized to the hole-doped case. We assume that the system is in
the weak scattering regime, i.e. $k_F l\gg 1$ where $k_F$ is the
Fermi wave vector and $l$ is the electron mean free path. It has
been found that $\sigma_{xy}^\text{II}$ vanishes~\cite{sini2007prb}
and the task gets reduced to the evaluation of
$\sigma_{xy}^\text{I}$ which is given by the following expression
\begin{equation}
\sigma_{xy}^\text{I}=\frac{e^2}{2\pi A}\text{Tr}\left\langle
\hat{v}_x
\hat{G}^R(\varepsilon_F)\hat{v}_y\hat{G}^A(\varepsilon_F)\right\rangle_c,
\end{equation}
where $\hat{G}^R$ and $\hat{G}^A$ are the retarded and advanced
Green's functions respectively, $\hat{v}_{x}$ and $\hat{v}_y$ are
the velocity operators, and the trace is taken over both momentum
and spin spaces. In weak scattering regime, the calculation is
performed perturbatively in the small parameter $1/(k_Fl)$.

In our model, we consider the scattering processes to be
quasi-elastic, hence the disorder lines in Feynman diagrams carry no
energy arguments. This serves as a good approximation for the
scattering by collective excitations such as phonons or magnons as
long as energy of collective excitation involved in the scattering
is much less than the Fermi energy. Since the typical energy scale
of excitations is $k_BT$, this condition is satisfied for
temperatures with $k_BT\ll \varepsilon_F$. Furthermore, for massless
excitation with a spectrum $\omega(k)=v_qk$ ($v_q$ is a constant
sound speed), quasi-elastic approximation is justified even at
higher temperatures if the quasi-particle speed $v_q$ is much less
than $v_F$, viz the band velocity at Fermi level. For massive
quasi-particle excitations, its validity can be justified if the
quasi-particle mass is much larger than the electron effective mass.

In the following, we calculate the Hall conductivity for each
individual class, or equivalently when one class of scattering is
dominant. The evaluation of the conductivity follows standard
procedures, and the relevant Feynman diagrams under self-consistent
non-cross approximation have been identified
before,~\cite{sini2007prb} so we do not elaborate here. The results
are listed below.

\subsection{Intrinsic Contribution}
The intrinsic contribution of AHE is a property purely of the
spin-orbit coupled band structure. It was first proposed by Karplus
and Luttinger,\cite{karp1954} and recently its connection with the
Berry phase of Bloch electrons is
established.\cite{jung2002,onod2002} It is now understood that
spin-orbit coupled bands usually possess effective magnetic fields
in momentum space known as Berry curvatures,\cite{sund1999} which
deflect carriers in the transverse directions. The intrinsic
contribution of anomalous Hall conductivity equals the integration
of Berry curvatures of all the occupied states. Because it does not
depend on scattering, intrinsic contribution is the same for all the
three universality classes. In Kubo-Streda formalism, intrinsic
contribution is the sum of the scattering-free part of
$\sigma_{xy}^\text{I}$ and $\sigma_{xy}^\text{II}$.

For electron doped case, we can separate the intrinsic Hall
conductivity $\sigma^\text{int}_{xy}$ into two parts,
\begin{equation}
\sigma^\text{int}_{xy}=\sigma^\text{int(v)}_{xy}+\sigma^\text{int(c)}_{xy},
\end{equation}
where $\sigma^\text{int(v)}_{xy}$ is the contribution from all the
completely occupied valence bands below the Fermi surface and
$\sigma^\text{int(c)}_{xy}$ is the contribution from the partially
filled conduction band where the Fermi surface lies in. The
contribution from completely filled bands
$\sigma^\text{int(v)}_{xy}$ must be a topologically quantized value
$Ce^2/(2\pi)$ with $C$ being an integer called the first Chern
number.\cite{naga2009} The lower band of the massive Dirac model has
a contribution of $-e^2/4\pi$. This is not a contradiction because
the Dirac band is not bounded. For any real physical system, the
evaluation of $C$ must go beyond the low energy effective model and
require complete information of the entire Fermi sea. On the
contrary, the contribution $\sigma^\text{int(c)}_{xy}$ from the
partially filled conduction band can be regarded as a Fermi surface
property\cite{hald2004} and is captured within the effective model,
\begin{equation}
\sigma^\text{int(c)}_{xy}=\frac{e^2}{4\pi}(1-\cos\theta_F),
\end{equation}
where $\theta_F$ is the spherical angle $\theta$ at the Fermi
surface when $k=k_F$.

\begin{figure}
  \centering
  \subfigure[]{
    \label{fig:a} 
    \includegraphics[width=1.3in]{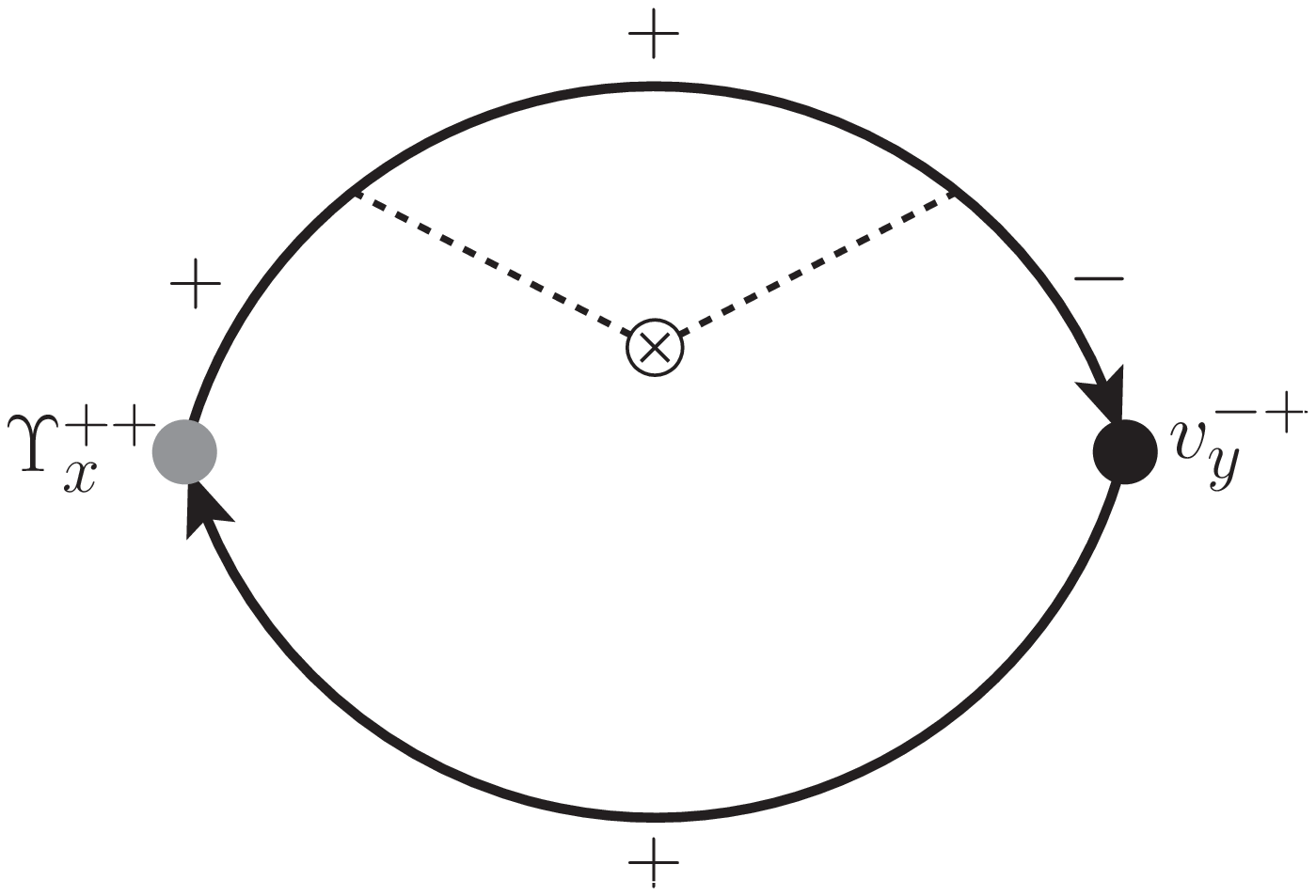}}
  \hspace{0.1in}
  \subfigure[]{
    \label{fig:b} 
    \includegraphics[width=1.3in]{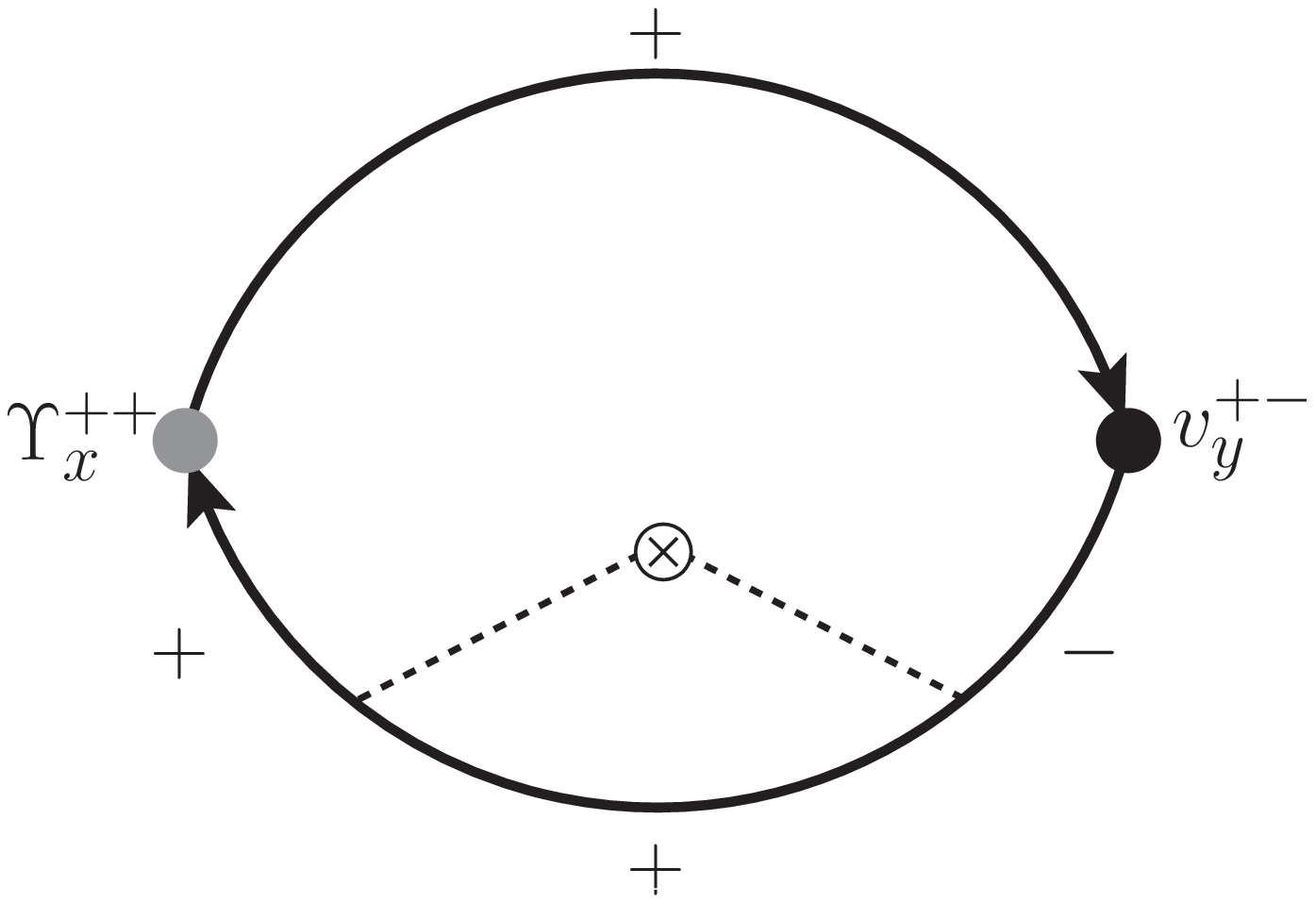}}
  \subfigure[]{
    \label{fig:c} 
    \includegraphics[width=1.3in]{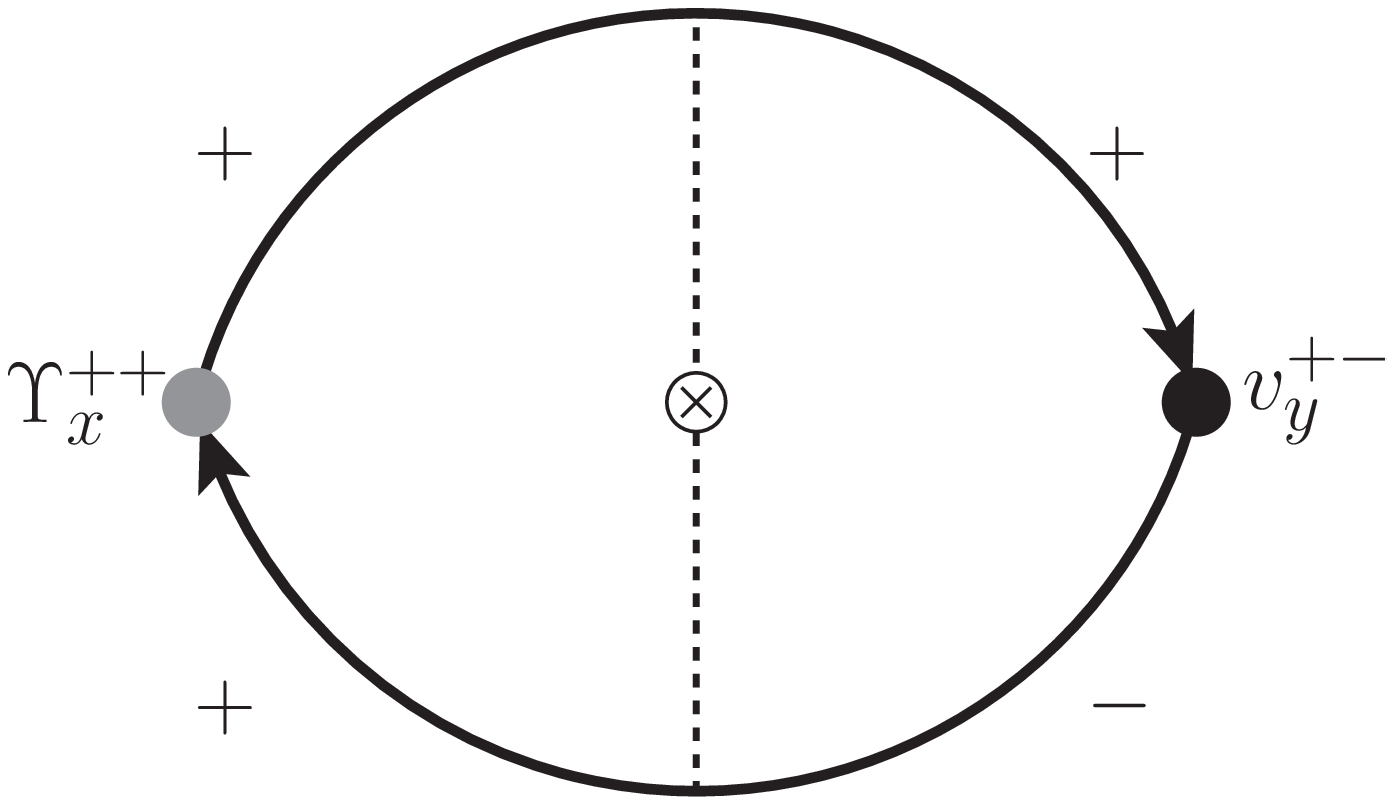}}
  \hspace{0.1in}
  \subfigure[]{
    \label{fig:d} 
    \includegraphics[width=1.3in]{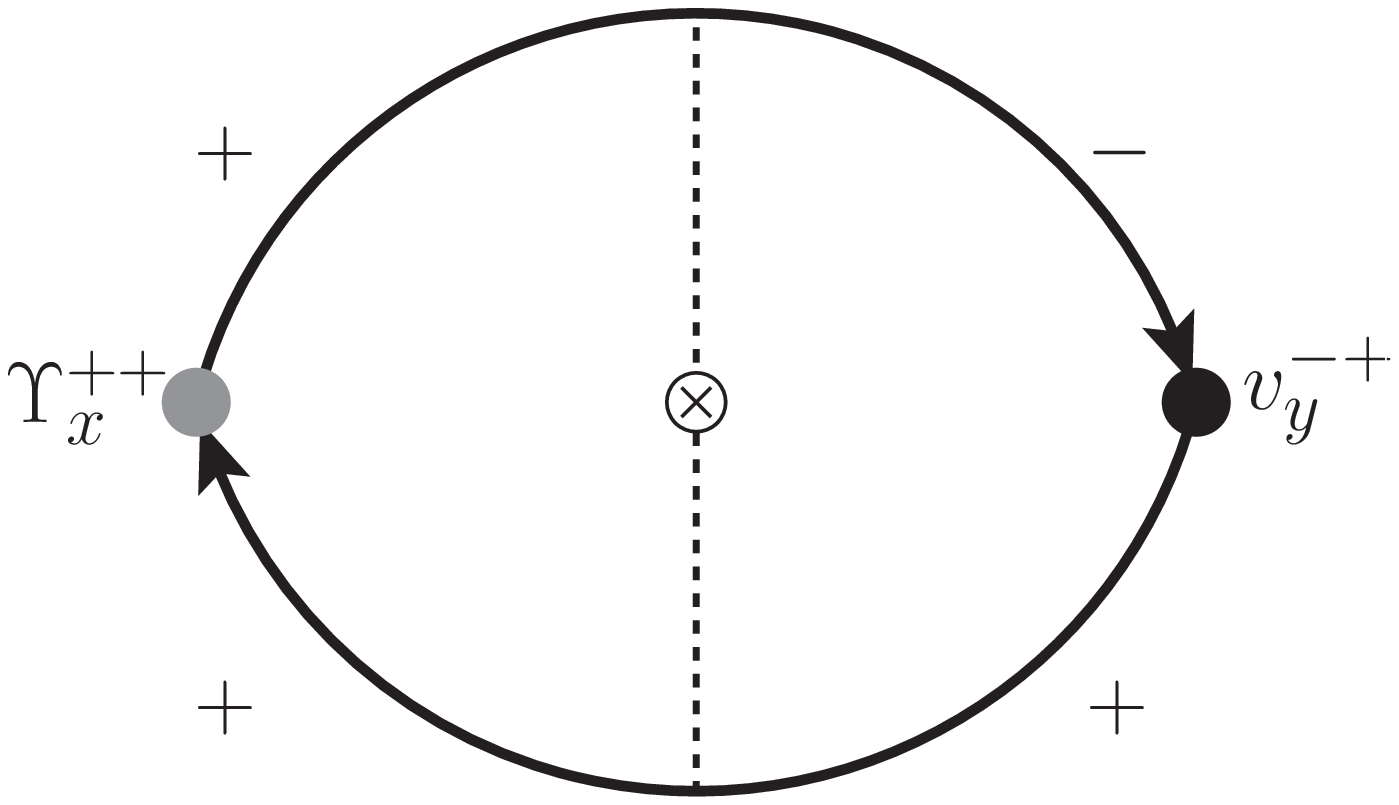}}
  \subfigure[]{
    \label{fig:e} 
    \includegraphics[width=1.4in]{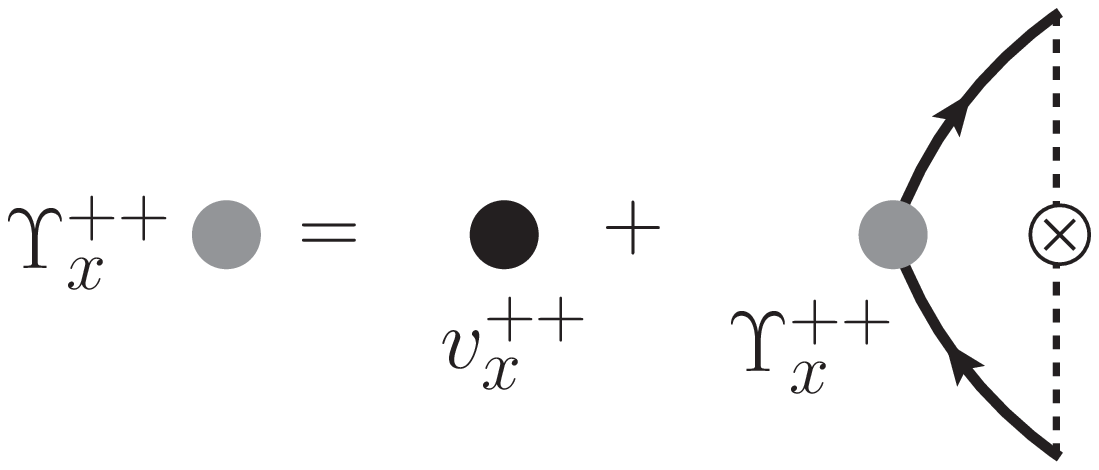}}
  \caption{(a)-(d) are the four conductivity diagrams of side jump that correspond to
  the distribution function correction. $+$ and $-$ represent the upper
  and lower band respectively.
  $\Upsilon$ stands for the renormalized velocity vertex
  which is dressed by a ladder like diagram as shown in (e).}
  \label{fig:adist} 
\end{figure}
\subsection{Side Jump Contribution}
Now let's focus on the side jump contribution which is the central
quantity we are interested in. For each individual class, it is
independent of disorder density $n_\text{dis}$ and scattering
strength $V^\text{o}$. It can be expressed in terms of $\theta_F$
and a set of scattering times defined on the Fermi surface. For
notational convenience, we define
\begin{equation}\label{stime}
\frac{1}{\tau_i}\equiv 2\pi
n_\text{dis}\int\frac{d^2\bm{k}'}{(2\pi)^2}\left|V_{\bm{k'k}}^\text{o}\right|^2
\cos^i(\phi-\phi')\delta(\varepsilon_F-\varepsilon_{\bm{k}'}^+),
\end{equation}
where
$V_{\bm{k'k}}^\text{o}=\langle\bm{k}'|V^\text{o}|\bm{k}\rangle$ is
the matrix element of the orbital part of the scattering potential
in momentum space and $i=0,1,2,\cdots$ is an integer.

In the semiclassical picture, the side jump we defined here consists
of three components: a contribution from the coordinate shift (the
original Berger's side jump), a contribution from a correction of
distribution function (called the anomalous distribution), and a
contribution from some higher order scattering processes (called the
intrinsic skew scattering). The first two components are shown to be
equal.~\cite{sini2007prb} In the Kubo-Streda approach,
Fig.~\ref{fig:adist} shows a set of diagrams that contributes to the
side jump in the chiral (eigenstate) basis. These correspond to the
contribution from the anomalous distribution function correction,
i.e. the second component above. The diagrams corresponding to the
contribution from coordinate shift can be obtained by simply
exchanging the subscripts $x$ and $y$ in Fig.~\ref{fig:adist} and
further making a $180^\circ$ rotation (i.e. exchanging $G^R$ and
$G^A$). The resulting contribution to Hall conductivity from these
two components for each scattering class is

\begin{figure*}
  \centering
  \subfigure[]{
    \label{fig:a} 
    \includegraphics[width=1.3in]{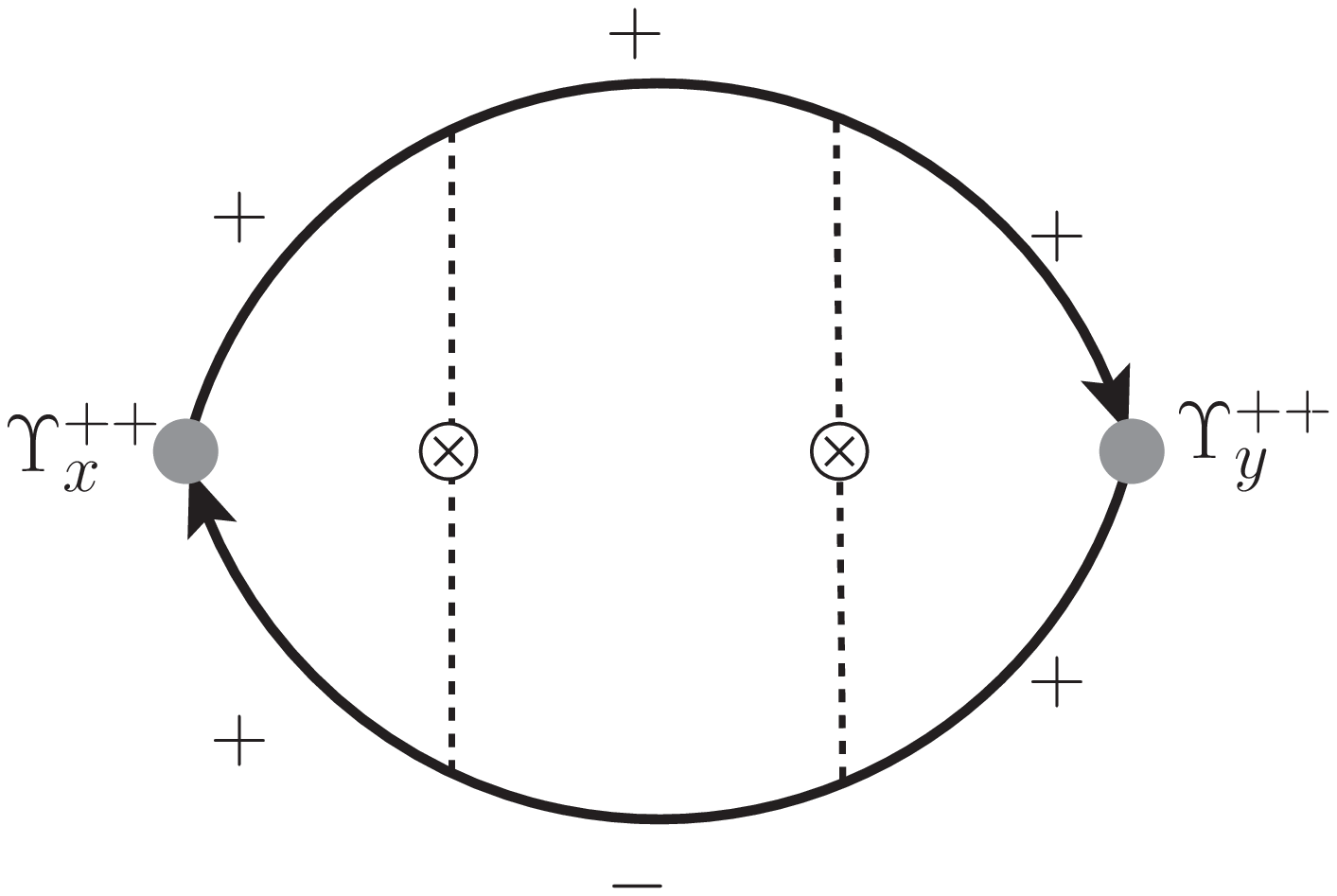}}
  \hspace{0.1in}
  \subfigure[]{
    \label{fig:b} 
    \includegraphics[width=1.3in]{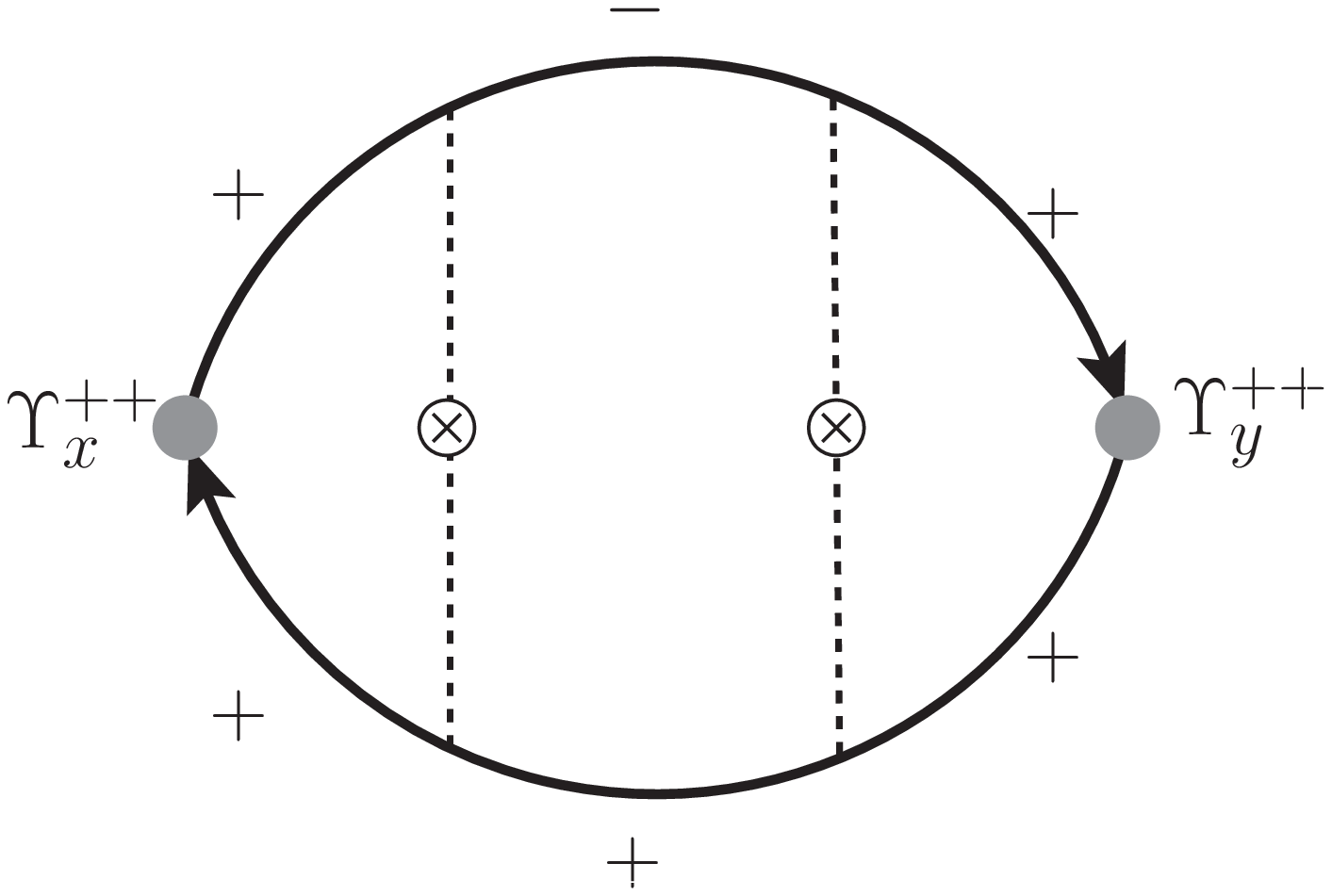}}
  \subfigure[]{
    \label{fig:c} 
    \includegraphics[width=1.3in]{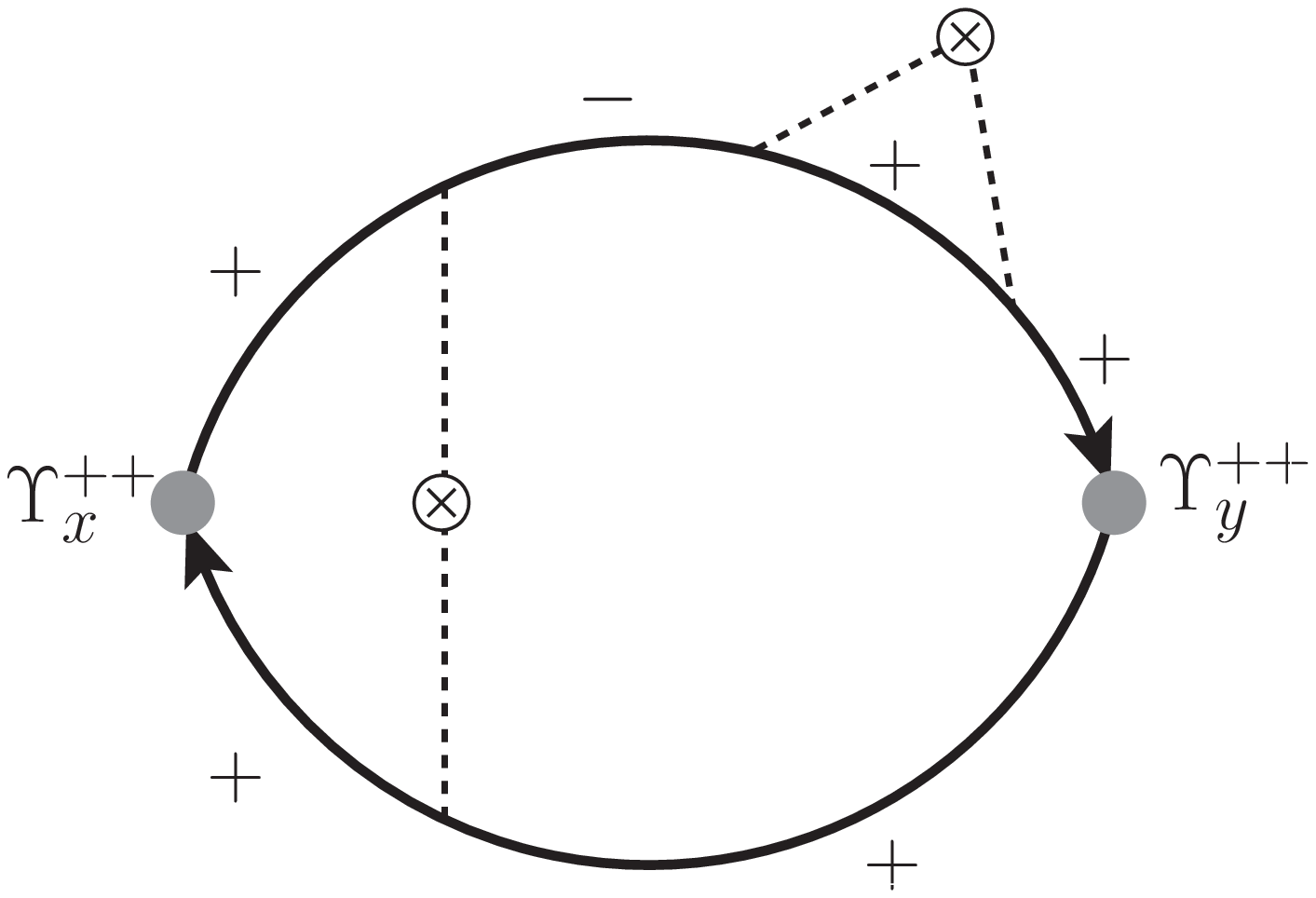}}
  \hspace{0.1in}
  \subfigure[]{
    \label{fig:d} 
    \includegraphics[width=1.3in]{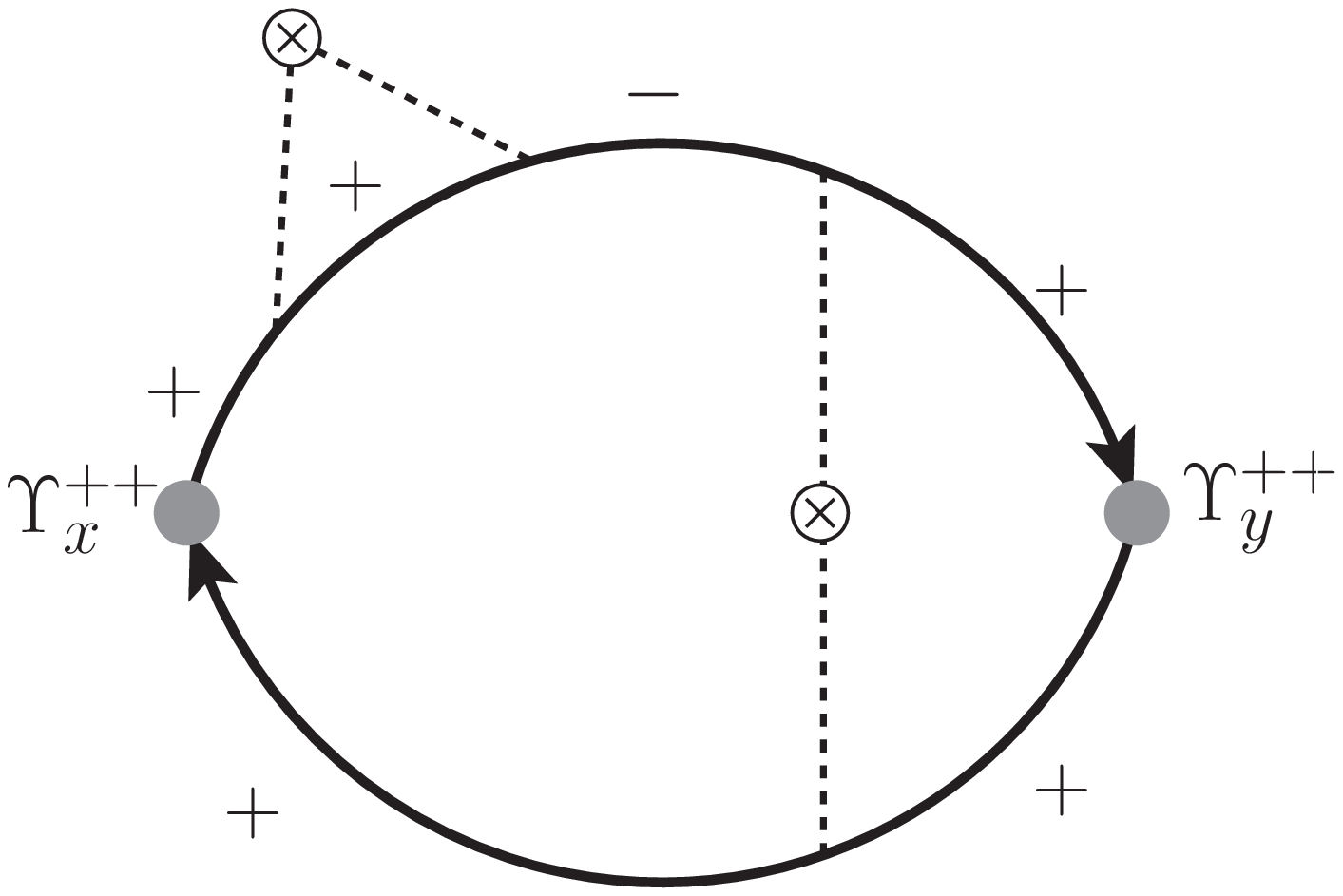}}
  \subfigure[]{
    \label{fig:e} 
    \includegraphics[width=1.3in]{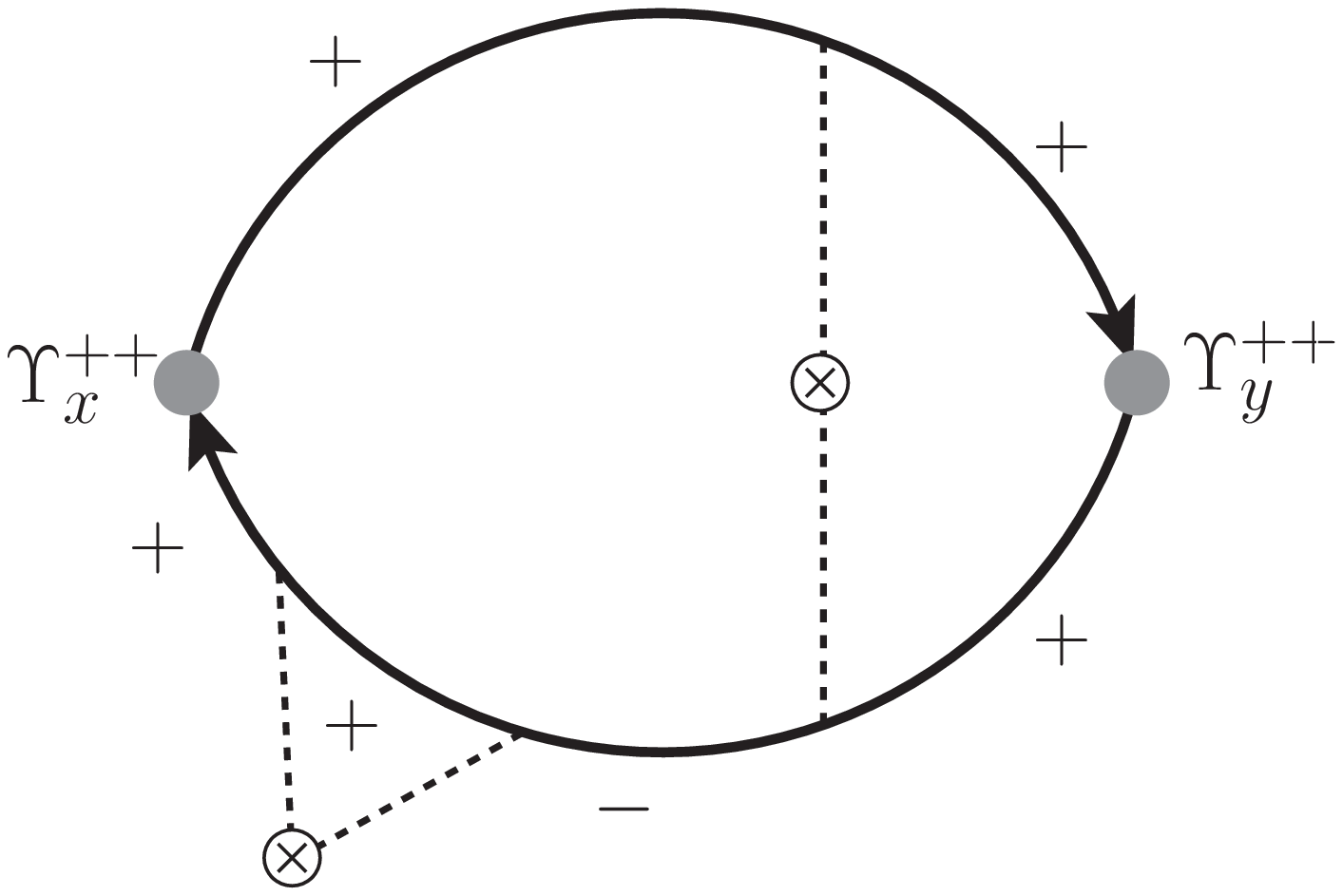}}
  \hspace{0.1in}
  \subfigure[]{
    \label{fig:f} 
    \includegraphics[width=1.3in]{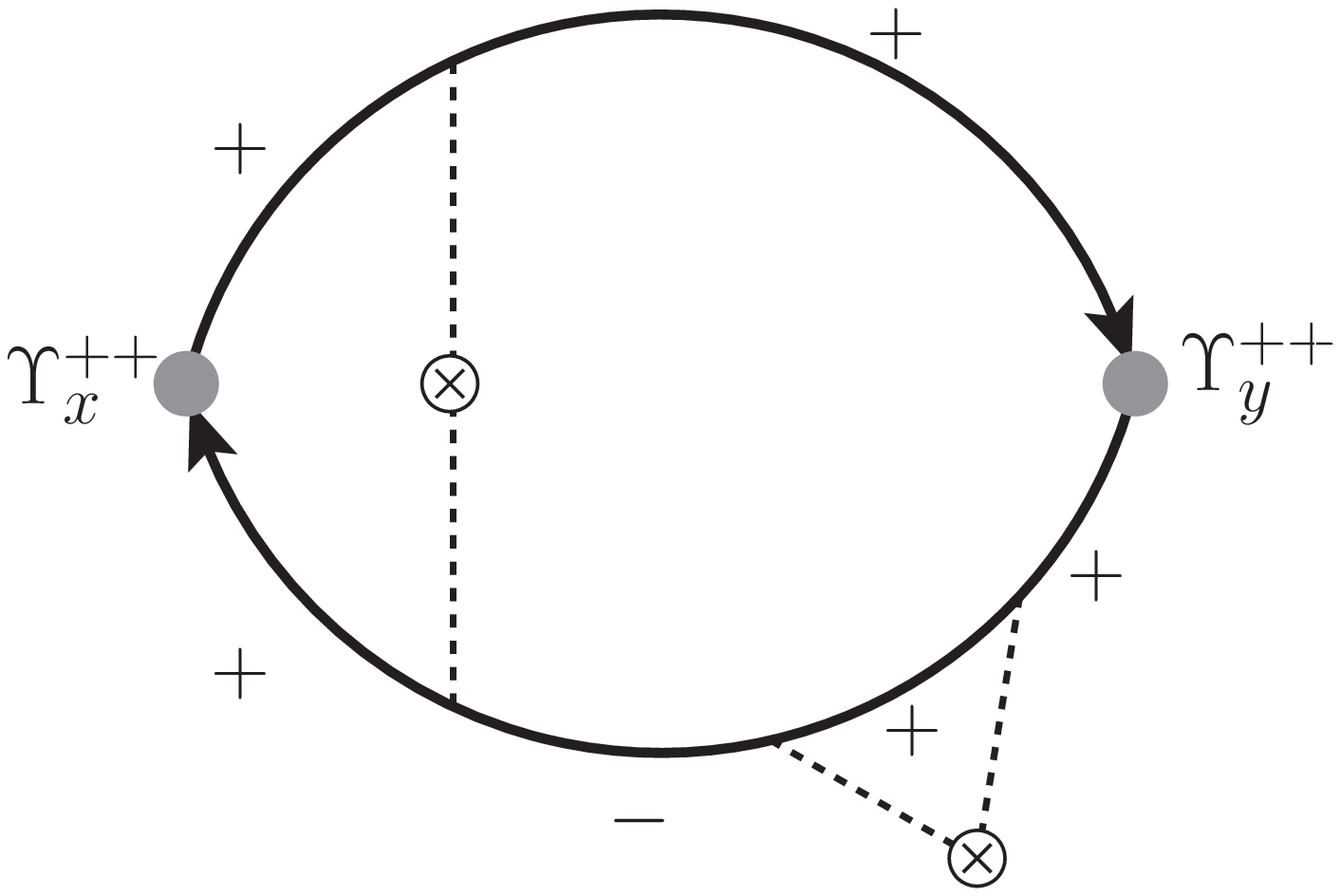}}
  \caption{The diagrams corresponding to the part of side jump from
  fourth order scattering process (intrinsic skew scattering). }
  \label{fig:isk} 
\end{figure*}
\begin{widetext}
\begin{equation}\begin{split}\label{sj}
&\text{Class A}:\qquad\sigma^\text{sj(a)}_{xy}=
-\frac{e^2}{2\pi}\frac{\sin^2\theta_F\cos\theta_F(\tau_0^{-1}-\tau_1^{-1})}
{(1+\cos^2\theta_F)\tau_0^{-1}-2\cos^2\theta_F\tau_1^{-1}-\sin^2\theta_F\tau_2^{-1}},\\
&\text{Class B}:\qquad\sigma^\text{sj(a)}_{xy}=0,\\
&\text{Class
C}:\qquad\sigma^\text{sj(a)}_{xy}=\frac{e^2}{4\pi}\cos\theta_F.
\end{split}
\end{equation}

We observe that different scattering class contributes very
differently to the Hall conductivity. In the diagramatic approach,
this difference originates from the different $\bm{k}$ dependence at
scattering vertices, which in turn results from their different spin
structures. It should be noted that the vanishing value of class B
is not a general feature but rather depends on the specific model we
considered here.\cite{classB}

The third component (intrinsic skew scattering) results from certain
fourth order scattering processes. The corresponding diagrams are
shown in Fig.~\ref{fig:isk} and the results are
\begin{equation}\begin{split}\label{isk}
&\text{Class A}:\qquad\sigma^\text{sj(b)}_{xy}=
-\frac{e^2}{4\pi}\frac{\sin^4\theta_F\cos\theta_F(\tau_0^{-1}-\tau_2^{-1})(\tau_0^{-1}-2\tau_1^{-1}+\tau_2^{-1})}
{\left[(1+\cos^2\theta_F)\tau_0^{-1}-2\cos^2\theta_F\tau_1^{-1}-\sin^2\theta_F\tau_2^{-1}\right]^2},\\
&\text{Class
B}:\qquad\sigma^\text{sj(b)}_{xy}=\frac{e^2}{4\pi}\frac{\sin^4\theta_F\cos\theta_F(\tau_0^{-1}-\tau_2^{-1})^2}
{\left[(1+\cos^2\theta_F)\tau_0^{-1}-2\tau_1^{-1}+\sin^2\theta_F\tau_2^{-1}\right]^2},\\
&\text{Class C}:\qquad\sigma^\text{sj(b)}_{xy}=0.
\end{split}
\end{equation}
\end{widetext}
The $\sigma^\text{sj(b)}_{xy}$ contribution vanishes for class C is
a very general result because each such diagram contains a factor of
the form $\int \sin\phi\text{d}\phi$ from the momentum integral at
the velocity vertex which suppresses the intrinsic skew scattering
process.

The total side jump contribution to anomalous Hall conductivity is
given by
$\sigma^\text{sj}_{xy}=\sigma^\text{sj(a)}_{xy}+\sigma^\text{sj(b)}_{xy}$.
It is clear that each class has a distinct side jump contribution.
Scattering rates with the same power appear in both nominator and
denominator of the expressions in Eqs.(\ref{sj},\ref{isk}), hence
the results are independent of disorder density and scattering
strength. The dependence on the band parameters such as $\Delta$ and
$\varepsilon_F$ are also different for different classes.
Furthermore, it should be noted that different classes can have side
jump contribution with different signs, as shown here between Class
A and the other two classes.

\subsection{Skew Scattering Contribution}
\begin{figure}
  \centering
  \subfigure[]{
    \label{fig:a} 
    \includegraphics[width=1.3in]{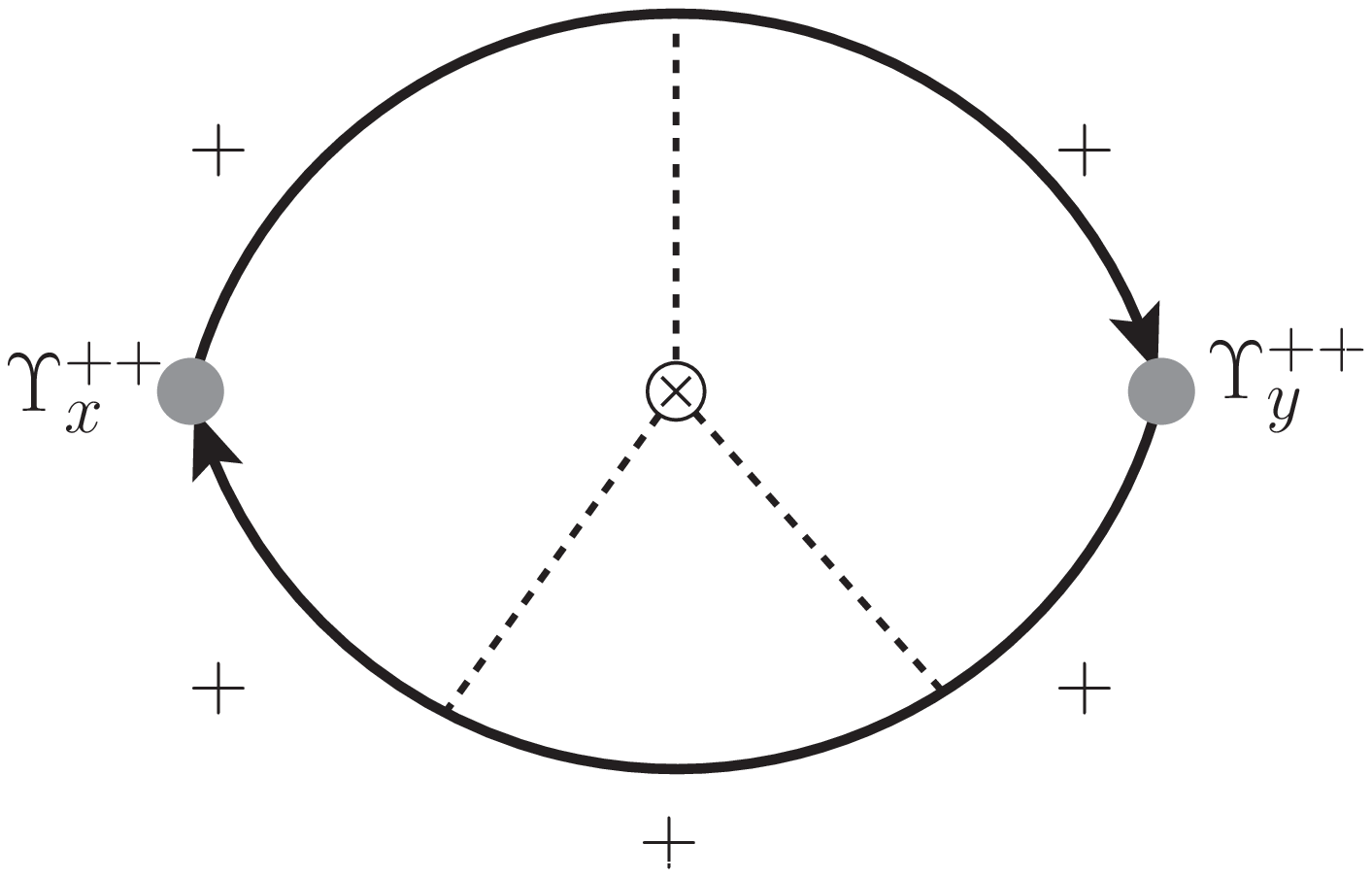}}
  \hspace{0.1in}
  \subfigure[]{
    \label{fig:b} 
    \includegraphics[width=1.3in]{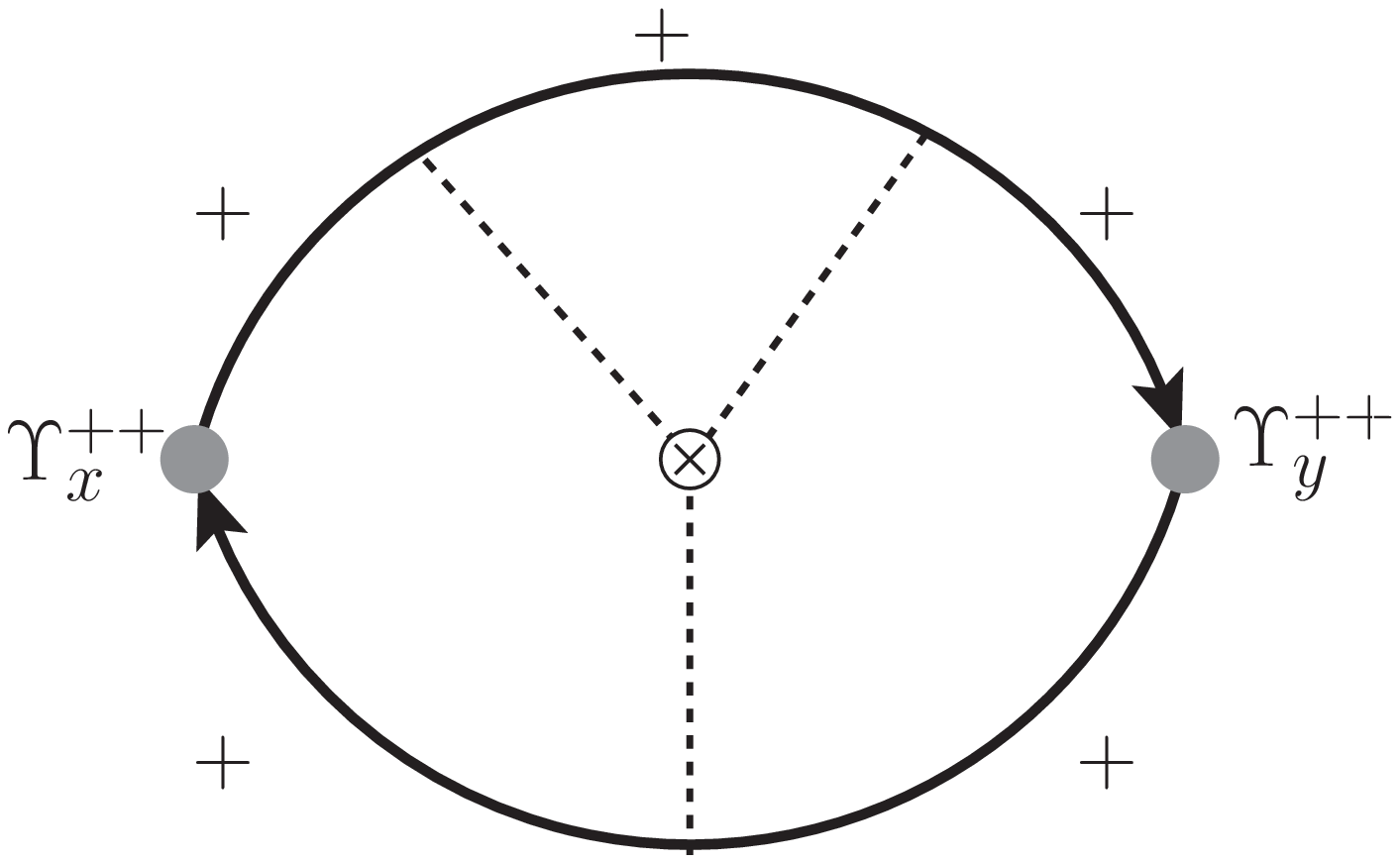}}
  \caption{The two diagrams corresponding to skew scattering contribution.}
  \label{fig:sk} 
\end{figure}

Although our focus is the side jump contribution, to be complete, we
also calculated the skew scattering contribution for each scattering
class. In the semiclassical picture, skew scattering contribution
comes from the asymmetric part of the scattering rates for higher
order scattering processes. The leading contribution is related to
the third order disorder correlation and has a $n_\text{dis}^{-1}$
dependence.\cite{fourth} The corresponding Feynman diagrams are
shown in Fig.~\ref{fig:sk} and the skew scattering contribution to
Hall conductivity for each class is given by

\begin{widetext}
\begin{equation}\begin{split}
&\text{Class A}:\qquad\sigma^\text{sk}_{xy}=
-\frac{e^2}{2\pi}\frac{\sin^4\theta_F\cos\theta_F\tau_\text{sk}^{-2}}
{\left[(1+\cos^2\theta_F)\tau_0^{-1}-2\cos^2\theta_F\tau_1^{-1}-\sin^2\theta_F\tau_2^{-1}\right]^2},\\
&\text{Class
B}:\qquad\sigma^\text{sk}_{xy}=\frac{e^2}{2\pi}\frac{\sin^4\theta_F\tau_\text{sk}^{-2}}
{\left[(1+\cos^2\theta_F)\tau_0^{-1}-2\tau_1^{-1}+\sin^2\theta_F\tau_2^{-1}\right]^2},\\
&\text{Class C}:\qquad\sigma^\text{sk}_{xy}=0.
\end{split}
\end{equation}
where
\begin{equation}\begin{split}
\frac{1}{\tau_\text{sk}^2}\equiv 2\pi
\varepsilon_Fn_\text{dis}\int\frac{d^2\bm{k}'}{(2\pi)^2}&\int\frac{d^2\bm{k}''}{(2\pi)^2}
\langle V_{\bm{kk'}}^\text{o}V_{\bm{k'k''}}^\text{o}V_{\bm{k''k}}^\text{o}\rangle_c\sin(\phi'-\phi)\\
&\cdot\left[
\sin(\phi-\phi'')+\sin(\phi''-\phi')+\sin(\phi'-\phi)\frac{}{}\right]
\cdot\delta(\varepsilon_F-\varepsilon^+_{\bm{k}'})
\delta(\varepsilon_F-\varepsilon^+_{\bm{k}''}).
\end{split}
\end{equation}
\end{widetext}
Note that the factor $1/\tau_\text{sk}^2$ is proportional to
$n_\text{dis}$, hence $\sigma^\text{sk}_{xy}$ is of order
$n_\text{dis}^{-1}$. Because the third order disorder correlation
vanishes for Class C as we discussed in Sec.II,
$1/\tau_\text{sk,C}^2=0$ and the skew scattering process is
forbidden for this type of disorder scattering. Therefore, we see an
important qualitative difference of the anomalous Hall response for
Class C and the other two classes: in the weak scattering regime,
the leading contribution for Class A and Class B is the skew
scattering of order $n^{-1}_\text{dis}$, but for Class C the leading
contribution is the intrinsic plus side jump which are of order
$n^{0}_\text{dis}$.

\subsection{Total Hall Conductivity of Order $n_\text{dis}^{0}$}
The above results are valid for disorder scattering with general
orbital part. If we consider the simple white noise (short range)
disorders, the results for Hall conductivity are greatly simplified.
In this case, we have for each class
\begin{equation}
\frac{1}{\tau_1}=0,\qquad \frac{1}{\tau_2}=\frac{1}{2\tau_0}.
\end{equation}
Then the total Hall conductivity can be written as
($\sigma_{xy}^\text{int(v)}$ is not included here, as
discussed in Sec.III.B):\\
\text{Class A:}
\begin{equation}\begin{split}\label{Adelta}
\sigma_{xy}=&\frac{e^2}{4\pi}(1-\cos\theta_F)-\frac{e^2}{\pi}\frac{\sin^2\theta_F\cos\theta_F}{1+3\cos^2\theta_F}
\\
&-\frac{3e^2}{4\pi}\frac{\sin^4\theta_F\cos\theta_F}{(1+3\cos^2\theta_F)^2}-
\frac{2e^2}{\pi}\frac{\sin^4\theta_F\cos\theta_F}{(1+3\cos^2\theta_F)^2}\frac{\tau_\text{sk}^{-2}}{\tau_0^{-2}}.
\end{split}
\end{equation}
\text{Class B:}
\begin{equation}\begin{split}
\sigma_{xy}=&\frac{e^2}{4\pi}(1-\cos\theta_F)+\frac{e^2}{4\pi}
\frac{\sin^4\theta_F\cos\theta_F}{(3+\cos^2\theta_F)^2}\\
&+
\frac{2e^2}{\pi}\frac{\sin^4\theta_F}{(3+\cos^2\theta_F)^2}\frac{\tau_\text{sk}^{-2}}{\tau_0^{-2}}.
\end{split}
\end{equation}
\text{Class C:}
\begin{equation}\begin{split}\label{Cdelta}
\sigma_{xy}=&\frac{e^2}{4\pi}.
\end{split}
\end{equation}
The expression in Eq.(\ref{Adelta}) recovers the result obtain by
Sinitsyn \emph{et al.}\cite{sini2007prb} Clearly, each universality
class has its distinct extrinsic Hall conductivity and different
functional dependence on the system parameters such as $\theta_F$.
Note that for class C, the side jump contribution cancels with the
part of intrinsic contribution which depends on the Fermi energy,
making the final result a constant and the value is the same as the
intrinsic contribution for a completely filled conduction band (i.e.
in the limit $\theta_F\rightarrow \pi/2$). We have checked that this
interesting cancelation occurs for a generic class of Hamiltonians
$\mathcal{H}=v k^n[\cos (n\phi_{\bm{k}})\sigma_x+\sin
(n\phi_{\bm{k}})\sigma_y]+\Delta\sigma_z$ where $n$ is an integer.

\begin{figure}
\includegraphics[width=0.8\columnwidth]{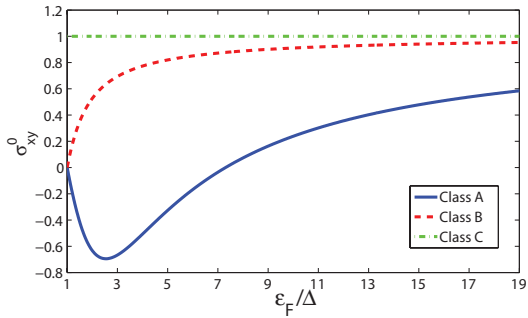}
\caption{\label{fig:sigma} (color online). $\sigma^0_{xy}$ plotted
as a function of the Fermi energy $\varepsilon_F$ for each of the
three universality classes. $\sigma^0_{xy}$ is measured in units of
$e^2/(4\pi)$, and $\varepsilon_F$ is measured in units of $\Delta$
which is half of the band gap.}
\end{figure}

Here we are most interested in the part of Hall conductivity
$\sigma_{xy}^0$ that is of order $n_\text{dis}^0$. This includes the
intrinsic contribution and the side jump contribution, i.e.
\begin{equation}
\sigma^0_{xy}=\sigma^\text{int}_{xy}+\sigma^\text{sj}_{xy}.
\end{equation}
In Fig.~\ref{fig:sigma}, for each scattering class (with white noise
spatial correlation), we plot $\sigma^0_{xy}$ as a function of Fermi
energy $\varepsilon_F$ for the massive Dirac model. Observe that
$\sigma^0_{xy}$ for class C takes a constant value $e^2/4\pi$ which
is independent of Fermi energy and the curves for both class A and
class B approach this constant value asymptotically as
$\varepsilon_F\rightarrow\infty$. For class A, the extrinsic
contribution has opposite sign as compared with the the intrinsic
contribution. Because of this sign difference, the Hall conductivity
for class A takes negative values for Fermi energies below
$\varepsilon_F\approx 7.3\Delta$. This behavior differs from that
for class B and class C whose extrinsic contributions have the same
sign as the intrinsic contribution, so their overall
$\sigma_{xy}^0$'s are positive. This shows that the Hall
conductivity sensitively depends on the class of disorder
scattering.

\subsection{Competition between Classes}
In the presence of two or more classes of scattering, there will be
a competition between them in the anomalous Hall response. The
resulting side jump contribution takes the following generic form:
\begin{equation}
\sigma^\text{sj}_{xy}=\frac{\sum_\alpha a_\alpha
\tau_\alpha^{-1}}{\sum_\alpha
b_\alpha\tau_\alpha^{-1}}+\frac{\sum_{\alpha\beta} c_{\alpha\beta}
\tau_\alpha^{-1}\tau_\beta^{-1}}{\sum_{\alpha\beta}
d_{\alpha\beta}\tau_\alpha^{-1}\tau_\beta^{-1}},
\end{equation}
where the $\tau_{\alpha(\beta)}$ is the scattering time defined for
each class of scattering involved, $a_\alpha$, $b_\alpha$,
$c_{\alpha\beta}$ and $d_{\alpha\beta}$ are the (disorder
independent) coefficients which depend only on system intrinsic
parameters such as $\theta_F$ in the present model.

As an example, let's consider the competition between Class A and
Class C. The calculation is tedious but straightforward. The
resulting total Hall conductivity can be expressed as
\begin{widetext}
\begin{equation}\label{sigAC}
\sigma_{xy}=\frac{e^2}{4\pi}(1-\cos\theta_F)-\frac{e^2}{\pi}\frac{\sin^2\theta_F\cos\theta_F(1-\zeta)}{
(1+3\cos^2\theta_F)+4\sin^2\theta_F\zeta}
-\frac{e^2}{\pi}\frac{\sin^4\theta_F\cos\theta_F(\frac{3}{4}-\zeta+2\eta)}{\left[
(1+3\cos^2\theta_F)+4\sin^2\theta_F\zeta\right]^2},
\end{equation}
\end{widetext}
where the parameter $\zeta$ defined as
$\zeta\equiv(\tau_{0C}^{-1}-\tau_{1C}^{-1})/(\tau_{0A}^{-1}-\tau_{1A}^{-1})$
is a measure of the relative disorder strength of the two classes,
and
$\eta\equiv\tau_{\text{sk},A}^{-2}/(\tau_{0A}^{-1}-\tau_{1A}^{-1})^2$
is a factor for skew scattering contribution from Class A, here
$\tau_{iA}$ stands for the scattering time $\tau_i$ defined in
Eq.(\ref{stime}) for Class A scattering and $\tau_{iC}$ is similarly
defined. The first term above is the intrinsic contribution, and the
remaining two terms (with $\eta=0$) are the side jump contribution.
Observe that in the limit $\zeta\rightarrow 0$ or $\zeta\rightarrow
+\infty$, Eq.(\ref{sigAC}) recovers our previous results in
Eq.(\ref{Adelta}) and Eq.(\ref{Cdelta}), and the value of Hall
conductivity varies continuously as $\zeta$ changes between these
two limits. This shows that the value of side jump is no longer
independent of disorder strength but can vary as a result of
competition between different scattering classes.

\begin{figure}
\includegraphics[width=0.8\columnwidth]{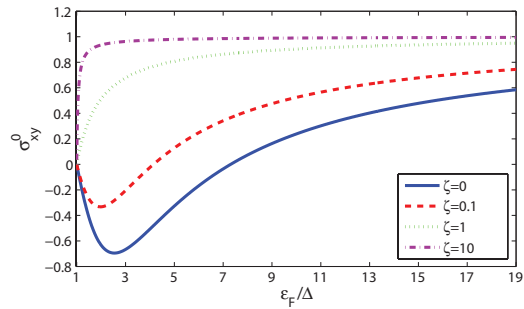}
\caption{\label{fig:sigvsefAC} (color online). $\sigma^0_{xy}$
plotted as a function of the Fermi energy $\varepsilon_F$ for fixed
values of $\zeta$. $\sigma^0_{xy}$ is measured in units of
$e^2/(4\pi)$, and $\varepsilon_F$ is measured in units of $\Delta$.}
\end{figure}

\begin{figure}
\includegraphics[width=0.8\columnwidth]{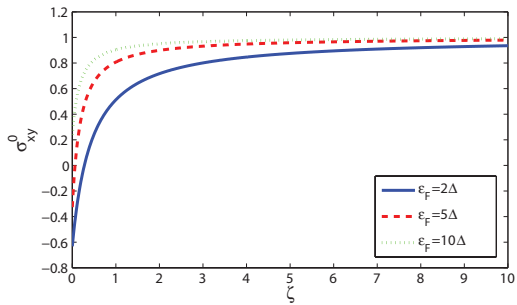}
\caption{\label{fig:sigvszetaAC} (color online). $\sigma^0_{xy}$
versus the ratio of scattering rates $\zeta$ for fixed values of
Fermi energy $\varepsilon_F$. $\sigma^0_{xy}$ is measured in units
of $e^2/(4\pi)$. The plot shows the crossover from Class A dominated
regime to Class C dominated regime as $\zeta$ increases.}
\end{figure}

In Fig.~\ref{fig:sigvsefAC}, we plot the Hall conductivity
$\sigma^0_{xy}$ (by setting $\eta=0$) as a function of the Fermi
energy for different values of $\zeta$. As $\zeta$ increases from
zero, the curve of $\sigma^0_{xy}$ is shifted upward from the Class
A dominated case due to the increasing contribution from Class C
scattering, and finally approaching the value $e^2/(4\pi)$ for the
Class C dominated case. This competition behavior is more clearly
seen in Fig.~\ref{fig:sigvszetaAC}, where $\sigma_{xy}^0$ is plotted
at three fixed Fermi energies as a function of $\zeta$. We see that
as $\zeta$ increases, $\sigma_{xy}^0$ increases monotonically. In
the energy range $\varepsilon<7.3\Delta$, there is a sign change of
$\sigma_{xy}^0$ during this crossover.

\section{Discussion and Conclusion}

As demonstrated in Sec.III, different scattering class has its own
distinct contributions to AHE. This suggests that for the study of
AHE in real materials, competing scattering processes belonging to
different classes need to be handled carefully.

In ferromagnetic materials, normal (non-magnetic) impurity
scattering and phonon scattering belong to Class A since they are
isotropic in spin space. Most of the previous studies on extrinsic
AHE are focused on this class of scattering and indeed it has been
found that the electron-phonon scattering has similar contribution
as normal impurity scattering (although there is no skew scattering
due to conservation of phonon population in steady
state),\cite{lyo1973} which is consistent with our theory. Magnetic
impurities with spin directions oriented along the average
magnetization is of Class B and they should generate a contribution
different from that of normal impurities. This has also been
observed in the study of dilute magnetic
semiconductors.~\cite{miha2008,nunn2008}

Scattering processes of Class C also exist. For example, magnetic
impurities with random in-plane magnetic orientation are of this
class. Moreover the scattering of electron by magnons also belongs
to Class C. To see this more clearly, let us consider the coupling
between conduction electron spin $\bm{\sigma}$ and the local spin
$\bm{S}$ (within an $s$-$d$ model approach),
\begin{equation}\begin{split}\label{exchange}
\hat{H}_\text{int}&=-J\int d\bm{r}\left[
\hat{\bm{\sigma}}(\bm{r})\cdot\bm{S}(\bm{r})\right]\\
&=-\frac{J}{2} \int
d\bm{r}\left(\hat{\sigma}_+S_-+\hat{\sigma}_-S_++2\hat{\sigma}_zS_z\right),
\end{split}
\end{equation}
where $J$ is the exchange coupling constant. The last term
$2\hat{\sigma}_zS_z$ describes the Zeeman splitting (which has
already been included in non-perturbed part of the Hamiltonian),
whereas the first two terms describe the scattering by magnetic
excitations. There is a transfer of spin angular momentum between
conduction electrons and local spins during this process, hence such
scattering is of Class C in our classification.

For real material samples that are studied experimentally, all the
three classes of scattering are present at finite temperature. At
low temperature, the scattering by normal impurity usually
dominates. With increasing temperature, electron-magnon scattering
becomes more important and can compete with normal impurity
scattering and phonon scattering. Especially for materials with a
high Debye temperature and a low Curie temperature, we can conceive
a situation in which Class A and Class C scattering compete as
depicted in Sec. III.F. Then the value of the side jump Hall
conductivity would flow between the two limiting values as a
function of temperature.

Finally we point out that the concept of ``spin" in our discussion
can be very general, corresponding to any discrete degrees of
freedom (sometimes called a ``pseudospin"). For example, in a
bipartite lattice (such as graphene), the sublattice degree of
freedom can be treated as pseudospin. Anomalous valley Hall
transport occurs in graphene when there is sublattice symmetry
breaking in the system.~\cite{xiao2007} For bilayer systems, it is
the layer index that plays the role of pseudospin. Then scattering
processes can be classified according to their effects on the
pseudospin. For example, inter-sublattice scattering in a bipartite
lattice or interlayer scattering in a bilayer system would both
belong to Class C. In general, our results indicate that a careful
analysis of various scattering processes according to their
pseudospin structures is indispensable for the study of AHE in these
systems.

In summary, we have shown that the extrinsic part of the anomalous
Hall conductivity has a strong dependence on the spin structure of
the disorder scattering. We propose three universality classes of
scattering according to their side jump contribution to the
anomalous Hall conductivity. Each class has its distinct value of
side jump. When two or more classes of scattering are competing, the
side jump contribution is determined by their relative disorder
strength. Various scattering processes in real physical systems can
be classified into these three classes. In particular, we
demonstrate that magnon scattering has distinct side jump
contribution from normal impurity scattering and phonon scattering
and the value of side jump contribution could change as system
control parameter (such as temperature) varies.
\\
\\
\\\emph{Acknowlegement} The authors thank N. A. Sinitsyn and J. Sinova
for helpful discussions. H. Pan was supported by NSFC (10974011). Y.
Yao was supported by NSFC (10974231) and the MOST Project of China
(2007CB925000). Q. Niu was supported by DOE (DE-FG02-02ER45958
Division of Materials Science and Engineering) and Texas Advanced
Research Program.


\begin{thebibliography}{99}
\bibitem{expl} The term which is proportional to $\langle
V_0V_z\rangle_c$ need not vanish under the symmetry constraint.
However, such scattering process is rearly physically relevant.
Therefore in the interest of transparency we do not consider this
term in the present work.
\bibitem{classB} For example, this contribution from Class B does not
vanish for the $k$-quadratic model
$\hat{\mathcal{H}}=v[(k_x^2-k_y^2)\hat{\sigma}_x+2k_xk_y\hat{\sigma}_y]+\Delta\hat{\sigma}_z$.
\bibitem{fourth} Contribution to skew scattering involving disorder
correlations higher than third order is conceivable. In our present
model of scattering, we do not consider such correlations.
\end{thebibliography}
\end{document}